\documentclass[a4paper,11pt,hyper]{JHEP3}
\usepackage{amsmath,latexsym,amssymb}
\usepackage{graphicx}
\usepackage{psfrag}
\usepackage[latin1]{inputenc}
\usepackage{subfigure}
\usepackage{nicefrac}
\usepackage{bbm}
\usepackage[stable]{footmisc}

\usepackage{comment}

\newcommand\nn{\nonumber}

\newcommand{\eq}[1]{\begin{equation}#1\end{equation}}
\newcommand{\spl}[1]{\begin{split}#1\end{split}}
\newcommand{\al}[1]{\begin{align}#1\end{align}}
\newcommand{\subeq}[1]{\begin{subequations}#1\end{subequations}}

\def\d{\text{d}}

\def\Re           {{\rm Re\hskip0.1em}}
\def\Im           {{\rm Im\hskip0.1em}}

\title{A landscape of non-supersymmetric AdS vacua on coset manifolds}

\author{Paul Koerber\footnote{Postdoctoral Fellow FWO -- Vlaanderen.}\\Instituut voor Theoretische Fysica, Katholieke Universiteit Leuven, Celestijnenlaan 200D, B-3001 Leuven, Belgium\\Email: \email{koerber} at \email{itf.fys.kuleuven.be}}
\author{Simon K\"ors\\Institut f\"ur Theoretische Physik, Universit\"at Heidelberg, Philosophenweg 16-19, D-69120 Heidelberg, Germany\\Email: \email{s.koers} at \email{thphys.uni-heidelberg.de}}

\abstract{We construct new families of non-supersymmetric sourceless type IIA
AdS$_4$ vacua on those coset manifolds that also admit supersymmetric solutions. We
investigate the spectrum
of left-invariant modes and find that most, but not all, of the vacua
are stable under these fluctuations. Generically, there are also no massless moduli.
}

\preprint{KUL-TF-09/28\\HD-THEP-09-31}

\begin{document}
\setcounter{footnote}{0}
\renewcommand{\thefootnote}{\arabic{footnote}}
\setcounter{section}{0}
\section{Introduction}

The reasons for studying AdS$_4$ vacua of type IIA supergravity are twofold:
first they are examples of flux compactifications away from the Calabi-Yau regime,
where all the moduli can be stabilized at the classical level. Secondly, they can
serve as a gravity dual in the AdS$_4$/CFT$_3$-correspondence, which became the
focus of attention due to recent progress in the understanding of the CFT-side
as a Chern-Simons-matter theory describing the world-volume of coinciding M2-branes \cite{abjm}.

It is much easier to find supersymmetric solutions of supergravity as the supersymmetry
conditions are simpler than the full equations of motion, while at the same time there are general theorems
stating that the former -- supplemented with the Bianchi identities of the form fields -- imply
the latter \cite{gauntlettM,lt,gauntlettIIB,integr}. Although special type IIA solutions
that came from the reduction of supersymmetric M-theory vacua were already known
(see e.g.~\cite{nilssonpope,sorokin1,sorokin2}), it was only in \cite{lt} that the supersymmetry conditions for type IIA vacua with SU(3)-structure
were first worked out in general. It was discovered that there are natural
solutions to these equations on the four coset manifolds $G/H$
that have a nearly-K\"ahler limit \cite{cveticnk1,cveticnk2,paltihouse,tomasiellocosets,cosets,effectivecosets} (solutions on other
manifolds can be found in e.g.~\cite{lt,dewolfe,acharya}).\footnote{For an early appearance of these coset manifolds in the string literature
see e.g.~\cite{lustcosets}.} To be precise these are the manifolds
SU(2)$\times$SU(2), $\frac{\text{G}_2}{\text{SU(3)}}$, $\frac{\text{Sp(2)}}{\text{S}(\text{U(2)}\times \text{U(1)})}$ and
$\frac{\text{SU(3)}}{\text{U(1)}\times \text{U(1)}}$.\footnote{See \cite{nkreview} for a review and a proof that these are the only homogeneous
manifolds admitting a nearly-K\"ahler geometry.}
These solutions are particularly simple in the sense that both the SU(3)-structure, which determines the metric, as well
as all the form fluxes can be expanded in terms of forms which are left-invariant under the action of the group $G$.
The supersymmetry equations of \cite{lt} then reduce to purely algebraic equations and can be explicitly solved. Nevertheless,
these solutions still have non-trivial geometric fluxes as opposed to the Calabi-Yau or torus orientifolds of \cite{dewolfe,acharya}.
Similarly to those papers it is possible to classically stabilize all left-invariant moduli \cite{effectivecosets}.
Inspired by the AdS$_4$/CFT$_3$ correspondence more complicated type IIA solutions have in the meantime been proposed. The solutions
have a more generic form for the supersymmetry generators, called SU(3)$\times$SU(3)-structure \cite{gualtieri}, and
are not left-invariant anymore \cite{tomasiellomassive,tomasiellomassive2,petriniSU3SU3,ltSU3SU3} (see also \cite{koerbercoisotropic}).
Supersymmetric AdS$_4$ vacua in type IIB with SU(2)-structure have also been studied in \cite{agataSU2,ltb,granascan,wraseSU2} and in particular
it has been shown in \cite{wraseSU2} that also in this setup classical moduli stabilization is possible.

At some point, however, supersymmetry has to be broken and we have to leave the safe haven of the supersymmetry conditions.
In this paper we construct new {\em non}-supersymmetric AdS$_4$ vacua without source terms.
This means that the more complicated equations of motion of supergravity should be tackled directly\footnote{Another route would be to find some alternative first-order equations,
which extend the supersymmetry conditions in that they still automatically imply the full equations of motion in certain non-supersymmetric cases,
see e.g.~\cite{gennonsusy,cassaninonsusy}.}. In order to simplify the equations we use a specific ansatz:
we start from a supersymmetric AdS$_4$ solution and scan for non-supersymmetric solutions with the {\em same} geometry (and thus SU(3)-structure),
but with {\em different} NSNS- and RR-fluxes. Moreover, we expand
these form fields in terms of the SU(3)-structure and its torsion classes. This may seem restrictive at first, but it works
for 11D supergravity, where solutions like this have been found and are known as Englert-type solutions \cite{englert,duffnonsusy1,duffnonsusy2} (see \cite{KKreview} for a review).
To be specific, for each supersymmetric M-theory solution of Freund-Rubin type (which means the M-theory four-form flux has only legs
along the external AdS$_4$ space, i.e.\ $F_4 = f \text{vol}_4$ where $f$ is called the Freund-Rubin parameter) it is possible
to construct a non-supersymmetric solution with the same internal geometry but with a different four-form flux. The modified four-form of the Englert solution
has then a non-zero internal part: $\hat{F}_4 \propto \eta^\dagger \gamma_{m_1m_2m_3m_4} \eta \, \d x^{m_1m_2m_3m_4}$, where $\eta$ is the 7D supersymmetry
generator, and a different Freund-Rubin parameter $f_E = -(2/3) f$. Also the Ricci scalar of the AdS$_4$ space, and thus the effective
4D cosmological constant, differs: $R_{\text{4D},E} = (5/6) R_{\text{4D}}$.
In type IIA with non-zero Romans mass (so that there is no lift to M-theory)
non-supersymmetric solutions of this form have been found as well: for the nearly-K\"ahler geometry in \cite{romansIIA,gennonsusy,cassred}
and for the K\"ahler-Einstein geometry in \cite{romansIIA,tomasiellomassive,dimitriosextrasol}. In this paper we show that this type of solutions
is not restricted to these limits and systematically scan for them. Applying our ansatz to the coset manifolds with nearly-K\"ahler limit,
mentioned above, we find that the most interesting manifolds are $\frac{\text{Sp(2)}}{\text{S}(\text{U(2)}\times \text{U(1)})}$
and $\frac{\text{SU(3)}}{\text{U(1)}\times \text{U(1)}}$, on which we find
several families of non-supersymmetric AdS$_4$ solutions. We also find some
non-supersymmetric solutions in regimes of the geometry that do not allow for a supersymmetric solution.

These non-supersymmetric solutions are not necessarily stable. For instance, it is known that if there is
more than one Killing spinor on the internal manifold (which holds in particular for $S^7$, the M-theory lift of
$\mathbb{CP}^3 = \frac{\text{Sp(2)}}{\text{S}(\text{U(2)}\times \text{U(1)})}$), the Englert-type
solution is unstable \cite{englertunstable}. We investigate stability of our solutions
against left-invariant fluctuations.
This means we calculate the spectrum of left-invariant modes, and check for each mode whether the mass-squared is above
the Breitenlohner-Freedman bound \cite{bf1,bf2}. This is not a complete stability analysis in that there
could still be non-left-invariant modes that are
unstable. We do believe it provides a good first indication. In particular, we find for the type IIA reduction of the Englert solution
on $S^7$ that the unstable mode of \cite{englertunstable} is among our left-invariant fluctuations and we find the exact same mass-squared.

These non-supersymmetric AdS$_4$ vacua are interesting, because, provided they are stable, they should
have a CFT-dual. For instance in \cite{tomasiellomassive} the CFT-dual for a non-supersymmetric
K\"ahler-Einstein solution on $\mathbb{CP}^3$ was proposed. Furthermore, for phenomenologically more realistic
vacua, supersymmetry-breaking is essential. Really, one would like to construct classical solutions with a dS$_4$-factor,
which are necessarily non-supersymmetric. Because of a series of no-go theorems -- from very general to more specific: \cite{malnun,kachrunogo,louisnogo,wrasenogo,koerbernogo} -- this is a very non-trivial task.
For papers nevertheless addressing this problem see \cite{silversteindS,vanrietds1,koerbernogo,vanrietds2,morenodS,wraseSU2}.
In this context the landscape of the non-supersymmetric AdS$_4$ vacua of this paper can be considered as a playground to gain experience before trying
to construct dS$_4$-vacua. In fact, in \cite{vanrietds2} an ansatz very
similar to the one used in this paper was proposed in order to construct dS$_4$-vacua. Applied to the coset manifolds above, it did however not
yield any solutions, in agreement with the no-go theorem of \cite{koerbernogo}.

In section \ref{sec:ansatz} we explain our ansatz in full detail, while in section \ref{sols} we present the explicit solutions
we found on the coset manifolds. In section \ref{stab} we analyse the stability against left-invariant fluctuations before
ending with some short conclusions. We provide an appendix with some
useful formulae involving SU(3)-structures and an appendix on our supergravity conventions.

The non-supersymmetric solutions of this paper appeared before in the second author's PhD thesis \cite{simonthesis}.

\section{Ansatz}
\label{sec:ansatz}

In this section we explain the ansatz for our non-supersymmetric solutions. The reader interested in the details might
want to check out our SU(3)-structure conventions in appendix \ref{asu3}, while towards the end of the section we need
the type II supergravity equations of motion outlined in appendix \ref{asugra}.

We start with a supersymmetric SU(3)-structure solution of type IIA supergravity. The SU(3)-structure
is defined by a real two-form $J$ and a complex decomposable three-form $\Omega$ satisfying \eqref{OmegaJ}.
Moreover, $J$ and $\Omega$ together determine the metric as in \eqref{su3metric}. In order for the solution
to preserve at least one supersymmetry ($N=1$) \cite{lt} one
finds that the warp factor $A$ and the dilaton $\Phi$ should be constant, the torsion classes $\mathcal{W}_1,\mathcal{W}_2$
purely imaginary and all other torsion classes zero (for the definition of the torsion classes see \eqref{torsionclasses}). This implies
\subeq{\label{dJdOm}\al{
\d J & = \frac{3}{2} W_1 \Re \Omega \, , \\
\d \Re \Omega & = 0 \, , \\
\label{imomder}
\d \Im \Omega & = W_1 J \wedge J + W_2 \wedge J \, ,
}}
where we defined $W_1 \equiv -i\mathcal{W}_1$ and $W_2\equiv-i\mathcal{W}_2$. The fluxes can then be expressed in terms of
$\Omega,J$ and the torsion classes and are given by
\subeq{\label{ansatz}\al{
e^{\Phi} \hat{F}_0 & = f_1 \, , \\
e^{\Phi} \hat{F}_2 & = f_2  \, J + f_3  \, \hat{W}_2 \, , \\
e^{\Phi} \hat{F}_4 & = f_4  \, J \wedge J + f_5  \, \hat{W}_2 \wedge J  \, , \\
e^{\Phi} \hat{F}_6 & = f_6 \, \text{vol}_6 \, , \\
H & = f_7 \, \Re \Omega \, ,
}}
where for the supersymmetric solution
\eq{\label{susyflux}\spl{
& f_1 = e^{\Phi} m \, , \quad f_2 = - \frac{W_1}{4} \, , \quad f_3 = - w_2 \, , \quad ~f_4 = \frac{3 \, e^{\Phi} m}{10} \, , \\
& f_5=0 \, , \quad ~~~~f_6 = \frac{9 \, W_1}{4} \, , \quad ~f_7= \frac{2 \, e^{\Phi} m}{5} \, .
}}
Using the duality relation $f=\tilde{F}_{0}= - \star_6 \hat{F}_6 = - e^{-\Phi} f_6$ (see \eqref{Fduality6d})
we find that $f_6$ is proportional to the Freund-Rubin parameter $f$, while $f_1$ is proportional to the
Romans mass $m$. Furthermore, we introduced here a normalized version of $W_2$, enabling us
later on to use \eqref{ansatz} as an ansatz for the fluxes also in the limit $W_2 \rightarrow 0$:
\eq{
\hat{W}_2 = \frac{W_2}{w_2} \, , \quad \text{with} \quad w_2 = \pm \sqrt{(W_2)^2} \, ,
}
where one can choose a convenient sign in the last expression.

The Bianchi identity for $\hat{F}_2$ imposes $\d W_2 \propto \Re \Omega$. Working out
the proportionality constant \cite{lt} we find
\eq{
\label{dW2}
\d W_2 = - \frac{1}{4} (W_2)^2 \Re \Omega \, .
}
Furthermore, using the values for the fluxes \eqref{susyflux} it fixes the Romans mass:
\eq{
\label{romanssusy}
e^{2\Phi} m^2 = \frac{5}{16} \left( 3 (W_1)^2 - 2(W_2)^2 \right) \, .
}

We now want to construct non-supersymmetric AdS solutions on the manifolds mentioned in the introduction with the {\em same} geometry
as in the supersymmetric solution, and thus the same SU(3)-structure $(J,\Omega)$, but with {\em different} fluxes. We make the ansatz
that the fluxes can still be expanded in terms of $J,\Omega$ and the torsion class $\hat{W}_2$ as in \eqref{ansatz}, but with different values
for the coefficients $f_i$. To this end we plug the ansatz for the geometry $(J,\Omega)$ --- eqs.~\eqref{dJdOm} --- and the ansatz for the fluxes --- eqs.~\eqref{ansatz} --- into the equations of motion \eqref{sugraeom} and solve for the $f_i$.
We will make one more assumption, namely that
\eq{
\label{W2W2}
\hat{W}_2 \wedge \hat{W}_2 = c \, J \wedge J + p \, \hat{W}_2 \wedge J \, ,
}
with $c,p$ some parameters. This is an extra constraint only for the  $\frac{\text{SU(3)}}{\text{U(1)}\times \text{U(1)}}$ coset and we will discuss its relaxation later.\footnote{With the ansatz \eqref{ansatz} the constraint is forced upon us.
Indeed, suppose that instead $\hat{W}_2 \wedge \hat{W}_2 = -1/6 \, J \wedge J + p \hat{W}_2 \wedge J + P \wedge J$,
where $P$ is a non-zero simple (1,1)-form independent of $\hat{W}_2$. We find then from the equation of motion
for $H$ and the internal part of the Einstein equation respectively $f_5 f_3=0$
and $(f_3)^2 - (f_5)^2 -(w_2)^2=0$. So the only possibility is then $f_5=0$ and $f_3 = \pm w_2$,
which leads in the end to the supersymmetric solution. They way out is to also include $P$ as an expansion form in \eqref{ansatz}.}
Wedging with $J$ we find then immediately $c=-1/6$.
Furthermore we need expressions for the Ricci scalar and tensor, which for a manifold with SU(3)-structure
can be expressed in terms of the torsion classes \cite{SU3riemann}. Taking into account that only $W_{1,2}$
are non-zero we find:
\subeq{\al{
R_{\text{6D}} & = \frac{15\, (W_1)^2 }{2}  - \frac{(W_2)^2}{2} \, , \\
R_{mn} & = \frac{1}{6 }g_{mn} R_{\text{6D}} + \frac{W_1}{4}  \, W_{2(m} \cdot J_{n)} + \frac{1}{2} \, \left[W_{2\,m}\cdot W_{2\,n}\right]_0 + \frac{1}{2} \Re \left[\d W_2|_{(2,1)\,m} \cdot \bar\Omega_n\right] \, , \label{riccitensor}
}}
where $(P)^2$ and $P_m \cdot P_n$ for a form $P$ are defined in \eqref{formsquared} and $|_0$ indicates
taking the traceless part. From eq.~\eqref{dW2} follows that for our purposes $\d W_2|_{2,1}=0$ so that the
last term in \eqref{riccitensor} vanishes. Moreover, using \eqref{W2W2} $\left[W_{2\,m}\cdot W_{2\,n}\right]_0$
can be expressed in terms of $W_{2(m} \cdot J_{n)}$.

Plugging the ansatz for the fluxes \eqref{ansatz} into the equations of motion \eqref{sugraeom} and
using eqs.~\eqref{dJdOm}, \eqref{dW2}, \eqref{simplehodge}, \eqref{W2W2}, \eqref{fluxcomp}, \eqref{Fduality6d} and \eqref{riccitensor}
we find:
\allowdisplaybreaks
\al{\label{eomconds}
\text{Bianchi} \, F_2 :~ 0&= \frac{3}{2} \, W_1 f_2 - \frac{1}{4} w_2 f_3 + f_1 f_ 7  \, ,\nn \\
\text{eom} \, F_4 :~ 0 &= 3 \, W_1 f_4 + \frac{1}{4} w_2 f_5 - f_6 f_7  \, ,\nn \\
\text{eom}\, H :~ 0&= 6 \, W_1 f_7 - 3 f_1 f_2 - 12 f_4 f_2 - 6 f_4 f_6  -f_3 f_5  \, , \nn\\
0 &= w_2 f_7 + f_1 f_3 + f_2 f_5 -2 f_3 f_4 - f_5 f_6  + p f_3 f_5  \, , \\
\text{dilaton eom} :~ 0 &= R_{\text{4D}} + R_{\text{6D}} - 2 f_7^2  \, ,\nn \\
\text{Einstein ext.} :~ 0&= R_{\text{4D}} + (f_1)^2 + 3 (f_2)^2 +12 (f_4)^2 + (f_6)^2 + (f_3)^2 +(f_5)^2  \, , \nn\\
\text{Einstein int.} :~ 0&= R_{\text{6D}} - 6 (f_7)^2 + \frac{1}{2} \left[3 (f_1)^2 + 3(f_2)^2 - 12 (f_4)^2 -3 (f_6)^2 + (f_3)^2 - (f_5)^2 \right] \, , \nn\\
0&= 4 (f_2 f_3+2 f_4 f_5) -  w_2 W_1  -  \, p \left[(f_3)^2-(f_5)^2-(w_2)^2 \right]  \, .\nn
}
In the equation of motion for $H$ we get separate conditions from the coefficients of $J\wedge J$ and $\hat{W}_2\wedge J$ respectively.
In the internal Einstein equation we find likewise a separate condition from the trace and the coefficient of $W_{2\, (m} \cdot J_{n)}$.
In the next section we find explicit solutions to these equations for the coset manifolds with nearly-K\"ahler limit, the stability
of which we investigate in section \ref{stab}.

\subsection*{Flipping signs}

The Einstein and dilaton equation are quadratic in the form fluxes and
thus insensitive to flipping the signs of these fluxes. Taking into account
also the flux equations of motion and Bianchi identities, we find that for
each solution to the supergravity equations, we automatically obtain new ones by making
the following sign flips:
\eq{\spl{\label{signflips}
H \rightarrow -H \, , \quad \hat{F}_0 \rightarrow -\hat{F}_0 \, , \quad \hat{F}_2 \rightarrow \hat{F}_2 \, , \quad \hat{F}_4 \rightarrow -\hat{F}_4 \, , \quad \hat{F}_6 \rightarrow \hat{F}_6 \, , \\
H \rightarrow -H \, , \quad \hat{F}_0 \rightarrow \hat{F}_0 \, , \quad \hat{F}_2 \rightarrow -\hat{F}_2 \, , \quad \hat{F}_4 \rightarrow \hat{F}_4 \, , \quad \hat{F}_6 \rightarrow -\hat{F}_6 \, , \\
H \rightarrow H \, , \quad \hat{F}_0 \rightarrow -\hat{F}_0 \, , \quad \hat{F}_2 \rightarrow -\hat{F}_2 \, , \quad \hat{F}_4 \rightarrow -\hat{F}_4 \, , \quad \hat{F}_6 \rightarrow -\hat{F}_6 \, .
}}
In particular, these sign flips will transform a supersymmetric solution into another supersymmetric solution (as can be verified
using the conditions \eqref{dJdOm},\eqref{susyflux} allowing for suitable sign flips of $J$, $\Re \Omega$ and $\Im \Omega$ compatible
with the metric). If some fluxes are zero, more sign flips are possible. For instance for $\hat{F}_0=\hat{F}_4=0$ we find the following
extra sign-flip, known as {\em skew-whiffing} in the M-theory compactification literature \cite{duffskewwhiffing} (see also the review \cite{KKreview})
\eq{
H \rightarrow \pm H \, , \quad \hat{F}_2 \rightarrow \hat{F}_2 \, , \quad \hat{F}_6 \rightarrow -\hat{F}_6 \, ,
}
which transforms a supersymmetric solution into a {\em non}-supersymmetric one. When discussing different solutions, we will from now on implicitly consider each solution together with its signed-flipped counterparts.

\section{Solutions}
\label{sols}

Let us now solve the equations obtained in the previous section for the coset manifolds that admit sourceless
supersymmetric solutions, namely $\frac{\text{G}_2}{\text{SU(3)}}$, SU(2)$\times$SU(2), $\frac{\text{Sp(2)}}{\text{S}(\text{U(2)}\times \text{U(1)})}$
and $\frac{\text{SU(3)}}{\text{U(1)}\times \text{U(1)}}$. For the supersymmetric solutions on these manifolds
we will use the conventions and presentation of \cite{cosets,effectivecosets}. For more details, including in particular
our choice of structure constants for the relevant algebras, we refer to these papers.

On a coset manifold $G/H$ one can define a coframe $e^m$ through the
decomposition of the Lie-valued one-form
$L^{-1}\d L = e^m \mathcal{K}_m + \omega^a \mathcal{H}_a$ in terms of the algebras of $G$ and $H$.
Here $L$ is a coset representative,
the $\mathcal{H}_a$ span the algebra of $H$ and the $\mathcal{K}_m$ span the complement
of this algebra within the algebra of $G$. The exterior derivative on the $e^m$
is then given in terms of the structure constants through the Maurer-Cartan relation. Furthermore,
the forms that are {\em left-invariant} under the action of $G$ are precisely those forms
that are constant in the basis spanned by $e^m$ and for which the exterior derivative is also
constant in this basis. For these forms the exterior derivative can then be expressed solely in terms
of the structure constants only involving the $\mathcal{K}_m$.
We refer to \cite{cosetreview1,cosetreview2} for a review on coset technology or
to the above papers for a quick explanation.

\subsection*{$\frac{\text{G}_2}{\text{SU(3)}}$ and SU(2)$\times$SU(2)}

We start from the supersymmetric nearly-K\"ahler solution on $\frac{\text{G}_2}{\text{SU(3)}}$.
The SU(3)-structure is given by
\eq{\spl{
J & = a (e^{12} - e^{34} + e^{56}) \, , \\
\Omega & = a^{3/2} \left[ (e^{245}-e^{236}-e^{146}-e^{135}) + i (e^{246}+e^{235}+e^{145}-e^{136}) \right] \, ,
}}
where $a$ is the overall scale.

Since this SU(3)-structure corresponds to a nearly-K\"ahler geometry the torsion class
$W_2$ is zero. Furthermore we find
\eq{
W_1 = - \frac{2}{\sqrt{3}} a^{-1/2} \, , \quad w_2=p=0 \, .
}
Plugging this into the equations \eqref{eomconds} we find exactly three solutions
for $(f_1,\ldots,f_7)$ (up to the sign flips \eqref{signflips}):
\eq{\label{G2sols}\spl{
& a^{-1/2} (\frac{\sqrt{5}}{2},\frac{1}{2\sqrt{3}},0,\frac{3}{4\sqrt{5}},0,- \frac{9}{2\sqrt{3}},\frac{1}{\sqrt{5}}) \, , \\
& a^{-1/2} (\sqrt{\frac{5}{3}},0,0,0,0,\frac{5}{\sqrt{3}},0) \, , \\
& a^{-1/2} (1,\frac{1}{\sqrt{3}},0,-\frac{1}{2},0,\sqrt{3},1) \, .
}}
The first is the supersymmetric solution,
while the last two are non-supersymmetric solutions, which were already found in \cite{romansIIA,gennonsusy,cassred}.
Truncating to the 4D effective theory it was shown in \cite{cassaninonsusy} that a generalization of this family of
solutions is quite universal as it appears in a large class of $N=2$ gauged supergravities.

On the SU(2)$\times$SU(2) manifold, requiring the same geometry as the supersymmetric solution
and not allowing for source terms will restrict us to the nearly-K\"ahler point. The analysis
is then basically the same as for $\frac{\text{G}_2}{\text{SU(3)}}$ above.

\subsection*{$\frac{\text{Sp(2)}}{\text{S}(\text{U(2)}\times \text{U(1)})}$}

The family of supersymmetric solutions on this manifold has, next to the overall scale, an extra parameter determining the shape
of the solutions. It is then possible to turn on the torsion class $W_2$ and venture away from the nearly-K\"ahler geometry.
This makes this class much richer and enables us this time to find new non-supersymmetric solutions.
The SU(3)-structure is given by \cite{tomasiellocosets,cosets,effectivecosets}
\eq{\spl{
J & = a (e^{12} + e^{34} - \sigma e^{56}) \, , \\
\Omega & = a^{3/2} \sigma^{1/2} \left[ (e^{245}-e^{236}-e^{146}-e^{135}) + i (e^{246}+e^{235}+e^{145}-e^{136}) \right] \, ,
}}
where $a$ is the overall scale and $\sigma$ is the shape parameter.
We find for the torsion classes and the parameter $p$:
\eq{\label{Sp2geom}\spl{
W_1 & = (a\sigma)^{-1/2} \, \frac{2+\sigma}{3} \, , \\
(W_2)^2 & = (a\sigma)^{-1} \, \frac{8(1-\sigma)^2}{3} \Rightarrow w_2 = (a\sigma)^{-1/2} \frac{2\sqrt{2}(1-\sigma)}{\sqrt{3}} \, , \\
\hat{W}_2 & = - \frac{1}{\sqrt{3}} \left( e^{12} + e^{34} + 2 \sigma e^{56} \right) \, , \\
p & = - \sqrt{2/3} \, .
}}
We easily read off that $\sigma=1$ corresponds to the nearly-K\"ahler geometry.
Note that even though $W_2 \rightarrow 0$ for $\sigma \rightarrow 1$, $\hat{W}_2$ is well-defined and non-zero in this limit so that
we can still use it as an expansion form for the fluxes. The points $\sigma=2$ and $\sigma=2/5$ are also special, since eq.~\eqref{romanssusy} then
implies that the supersymmetric solution has zero Romans mass and, in particular, can be lifted to M-theory. Moreover, these
are the endpoints of the interval where supersymmetric solutions exist (since outside this interval we would find from eq.~\eqref{romanssusy}
that $m^2<0$). They are indicated as vertical dashed lines in the plots.

\begin{figure}
\centering
\includegraphics[width=14cm]{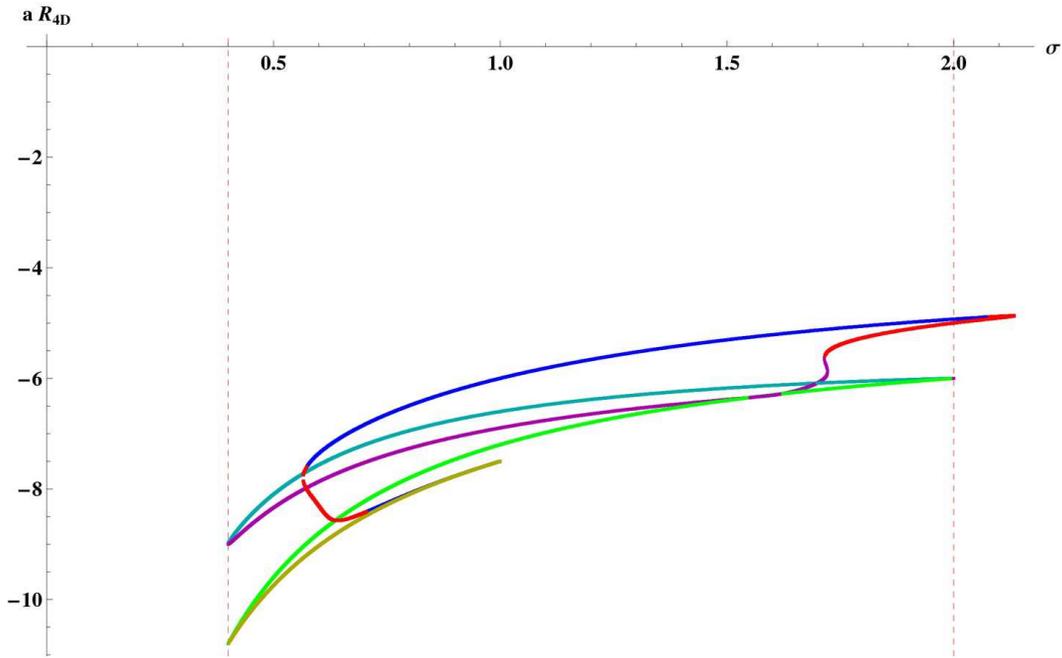}
\caption{$\frac{\text{Sp(2)}}{\text{S}(\text{U(2)}\times \text{U(1)})}$-model: plot of $a R_{\text{4D}}$ for the supersymmetric solutions (light green) and the new non-supersymmetric solutions
(other colors) in terms of the shape parameter $\sigma$. Unstable solutions are indicated in red.}
\label{plotR4dSp2}
\end{figure}

\begin{figure}
\centering
\subfigure[Plot of $a^{1/2} f_1$ (Romans mass)]{
\includegraphics[width=7cm]{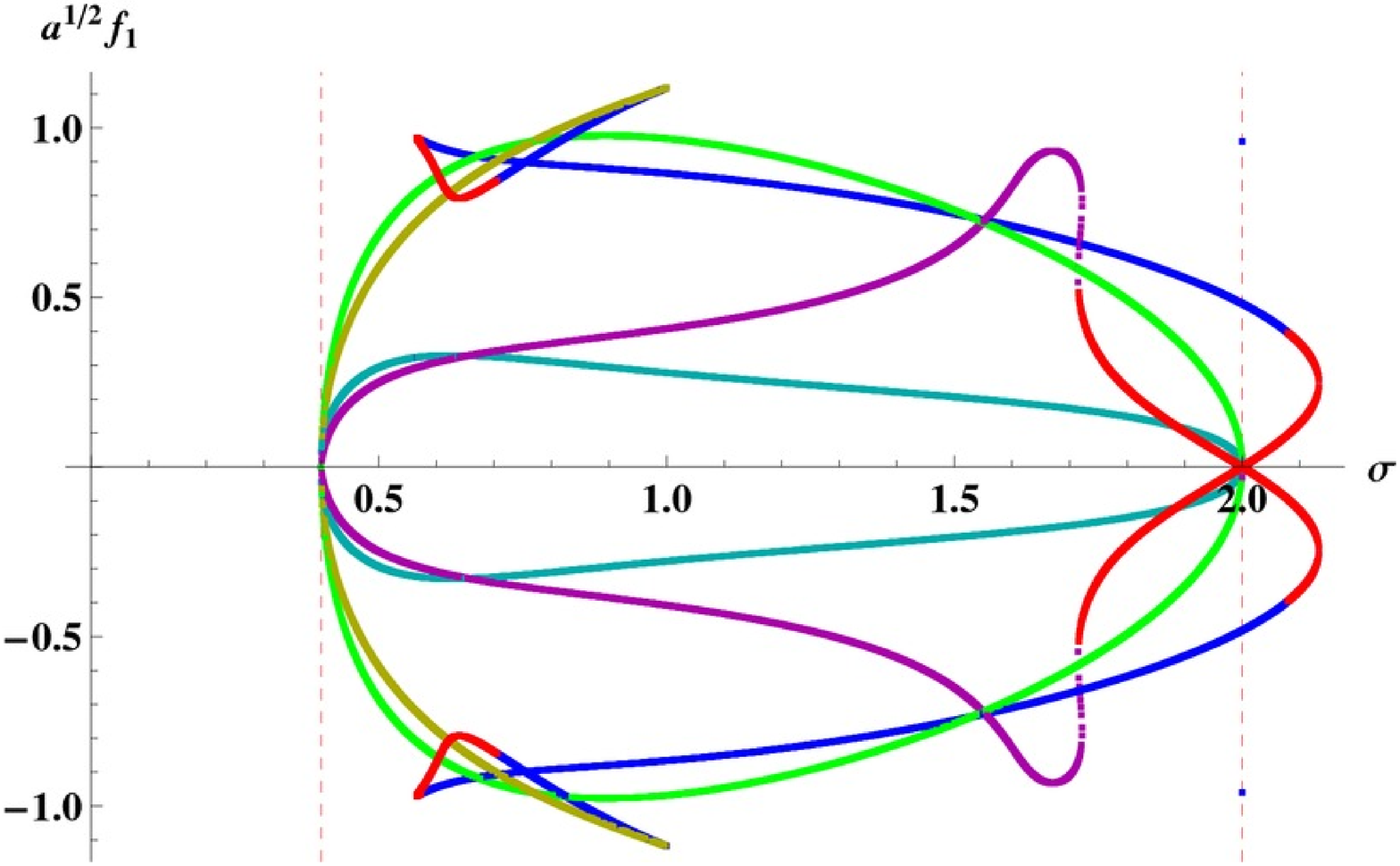}
}
\hspace{0.2cm}
\subfigure[Plot of $a^{1/2} f_2$ ($J$-part of $\hat{F}_2$)]{
\includegraphics[width=7cm]{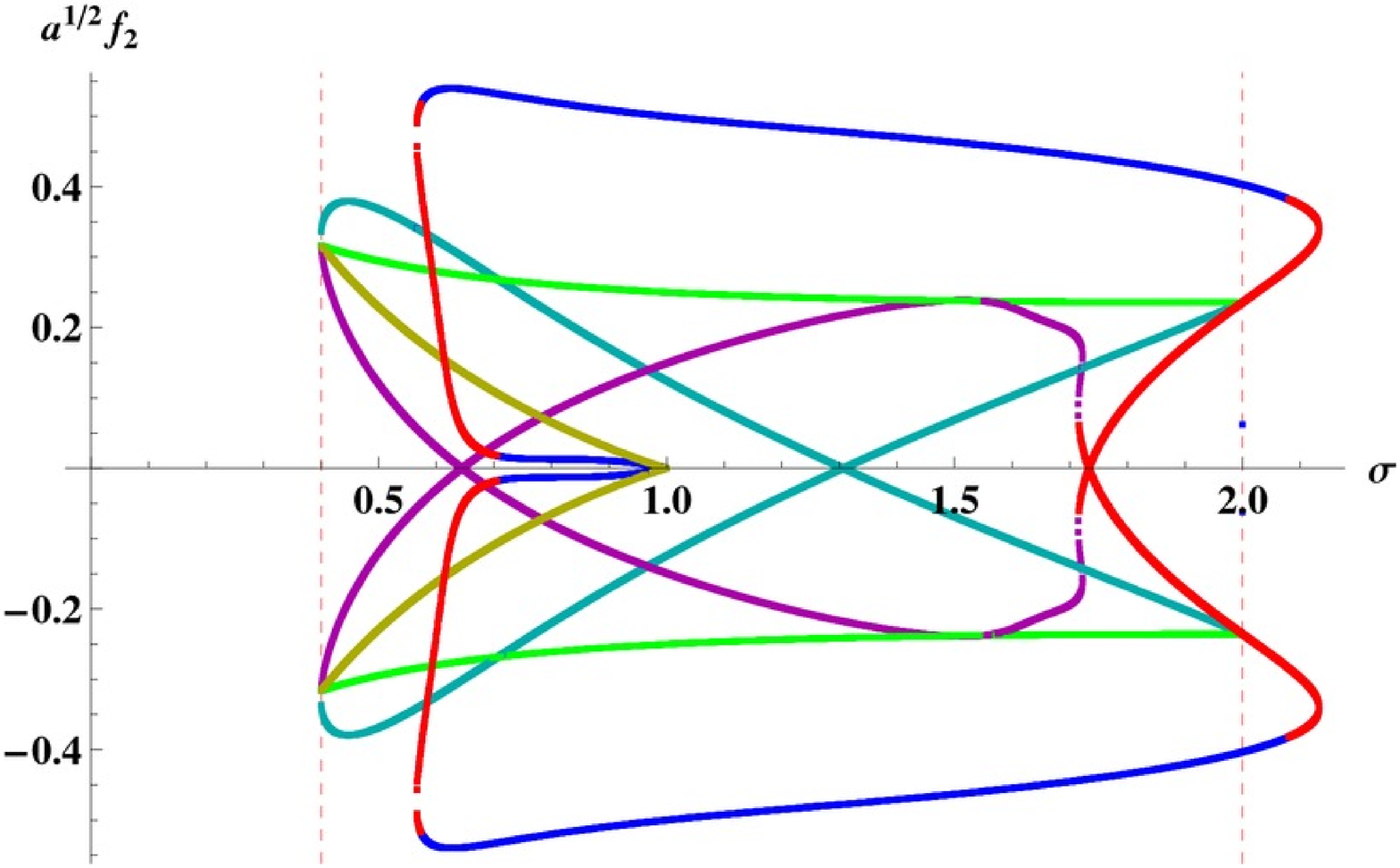}
}\\
\subfigure[Plot of $a^{1/2} f_3$ ($\hat{W}_2$-part of $\hat{F}_2$)]{
\includegraphics[width=7cm]{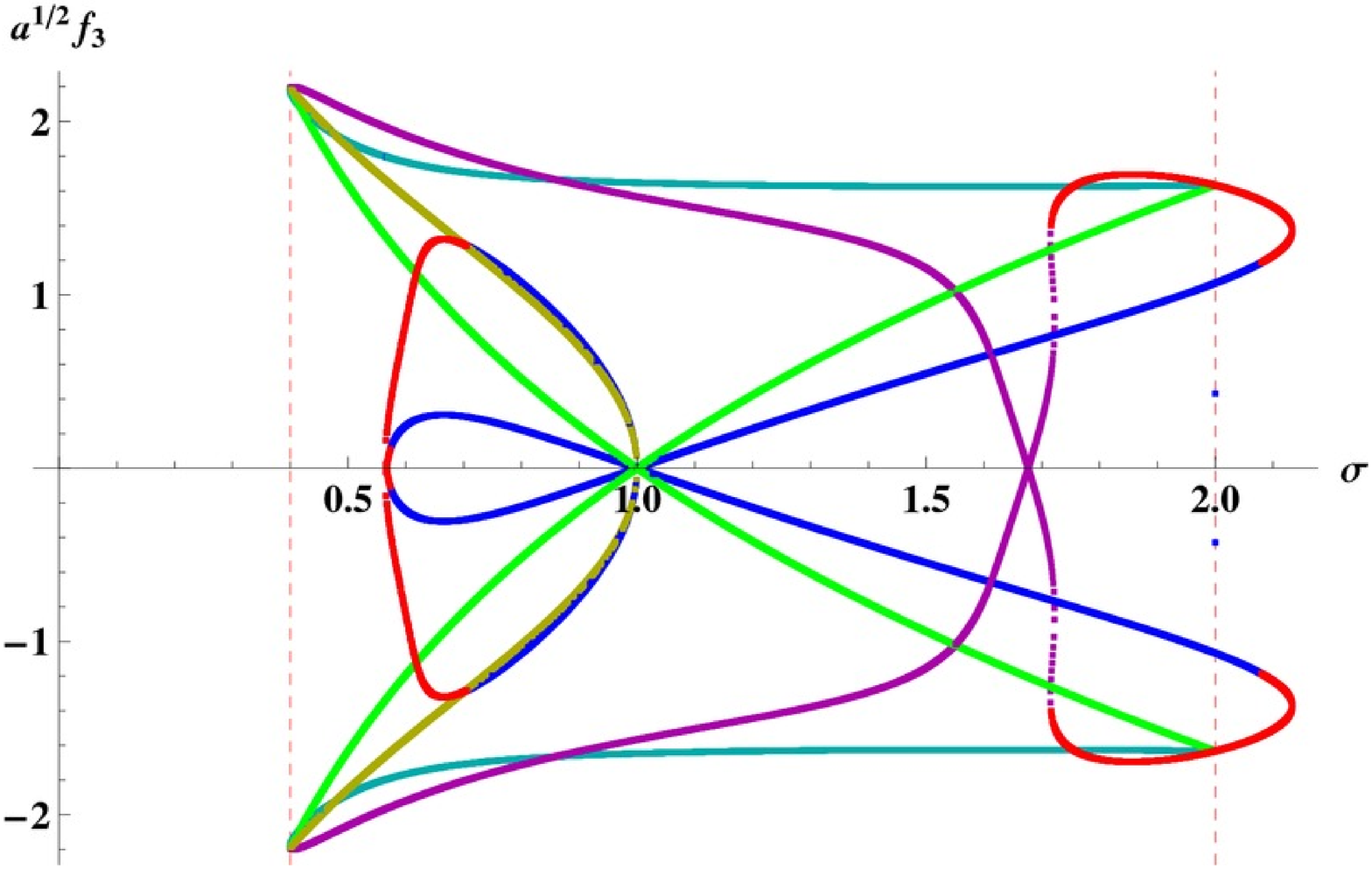}
}\hspace{0.2cm}
\subfigure[Plot of $a^{1/2} f_4$ ($J\wedge J$-part of $\hat{F}_4$)]{
\includegraphics[width=7cm]{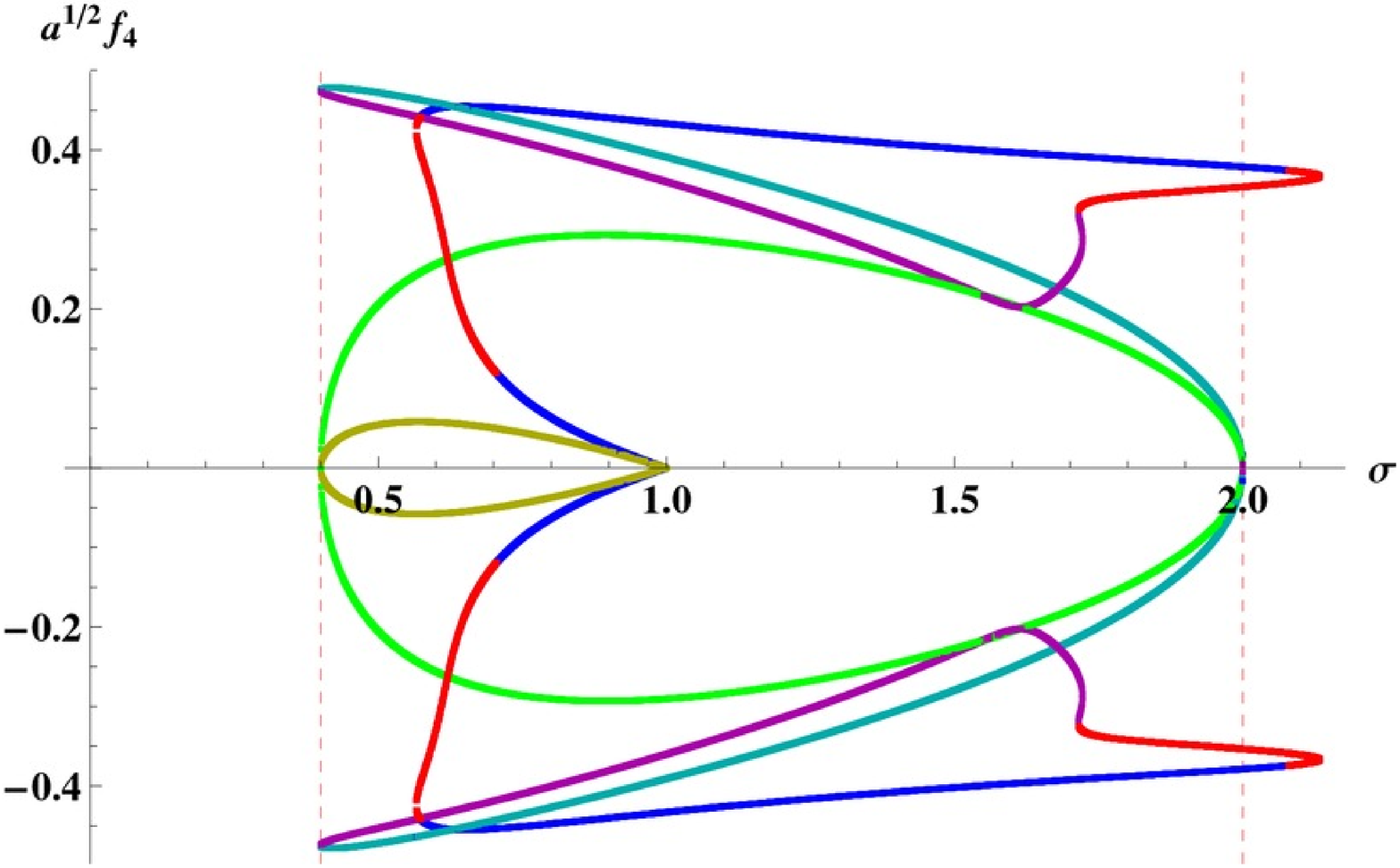}
}\\
\subfigure[Plot of $a^{1/2} f_5$ ($J\wedge \hat{W}_2$-part of $\hat{F}_4$)]{
\includegraphics[width=7cm]{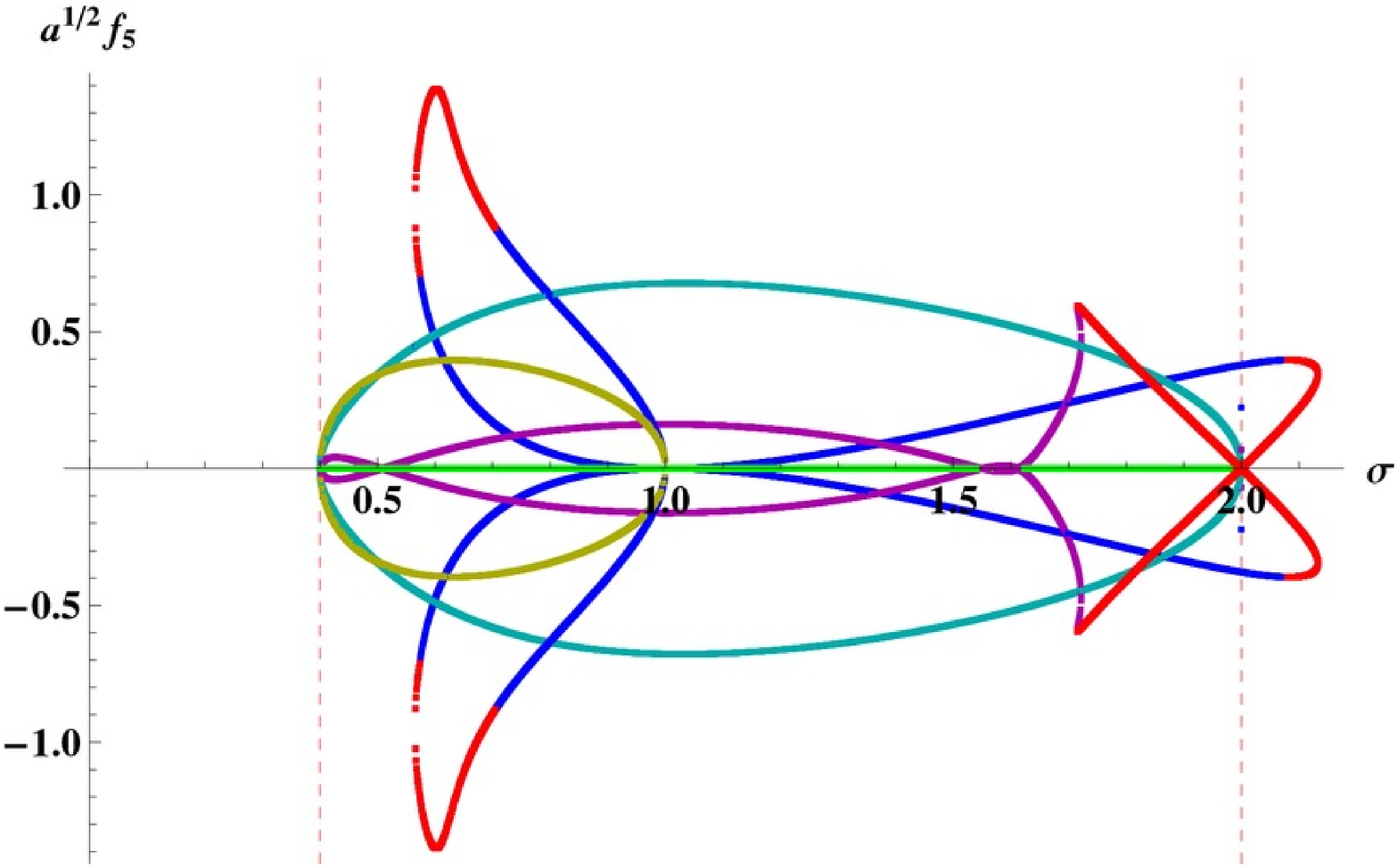}
}\hspace{0.2cm}
\subfigure[Plot of $a^{1/2} f_6$ (Freund-Rubin parameter)]{
\includegraphics[width=7cm]{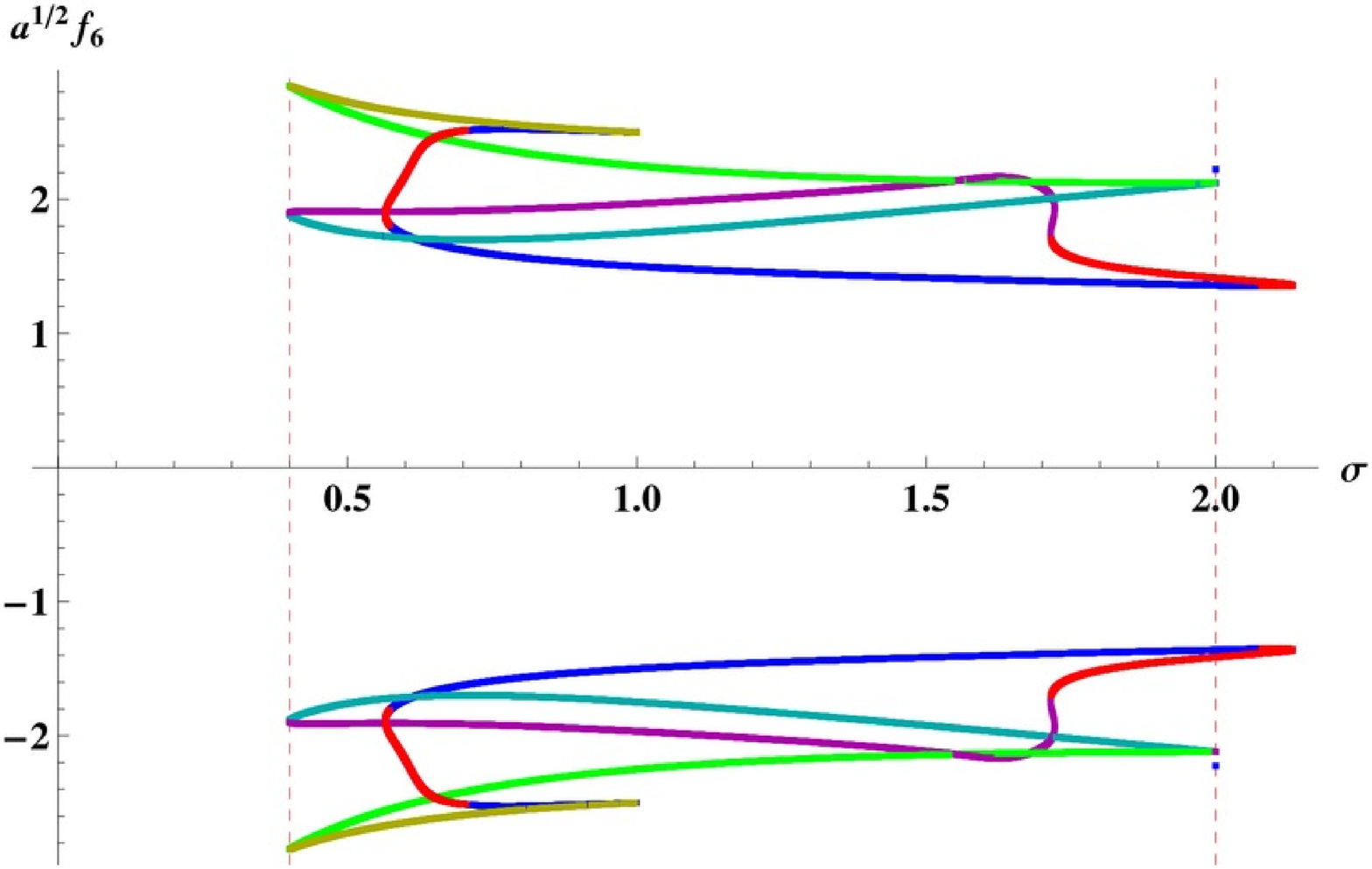}
}\\
\subfigure[Plot of $a^{1/2} f_7$ ($\Re\Omega$ part of $H$)]{
\includegraphics[width=7cm]{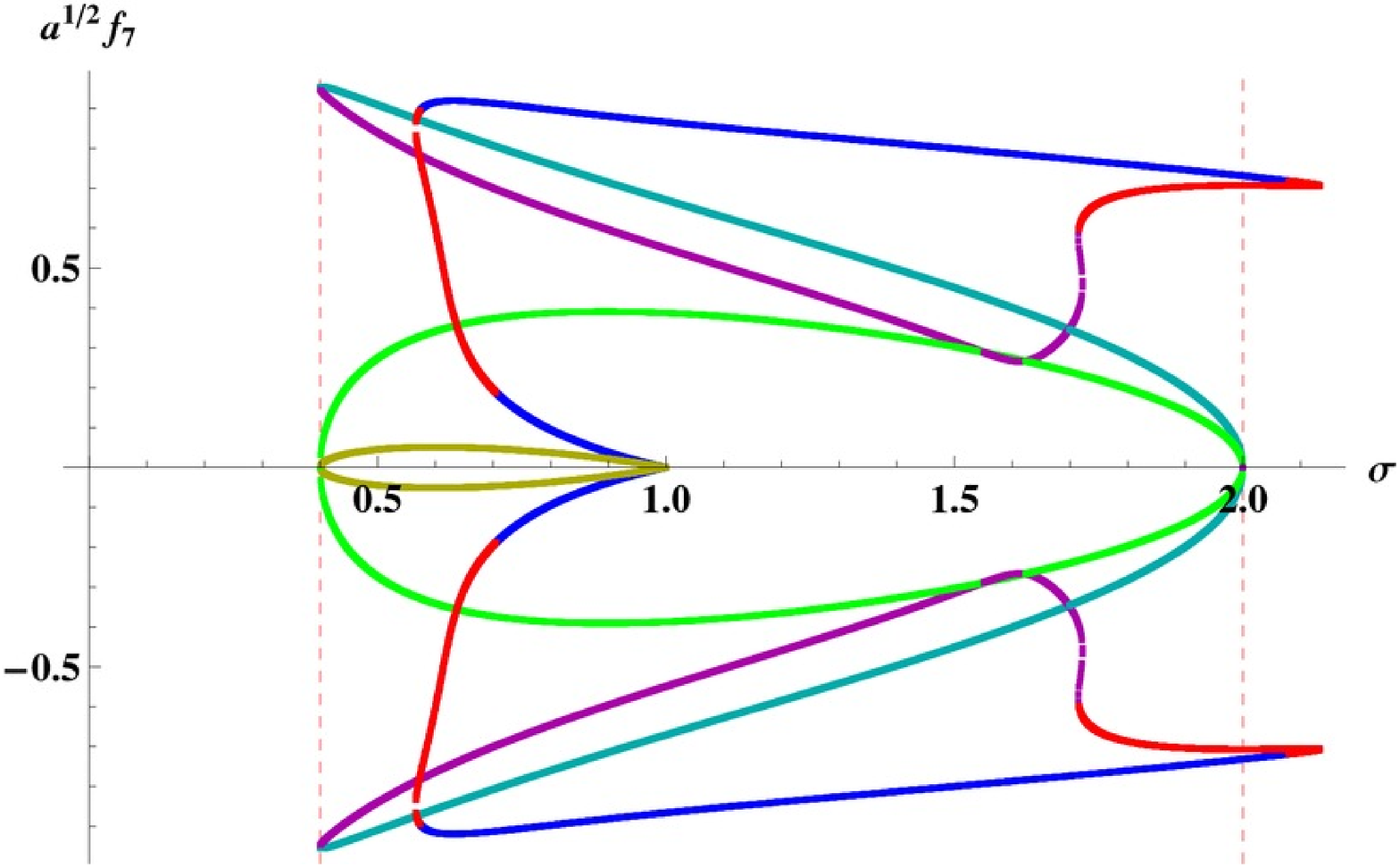}
}
\caption{Plots of the solutions on the coset
$\frac{\text{Sp(2)}}{\text{S}(\text{U(2)}\times \text{U(1)})}$. Different colors
indicate different solutions. Unstable solutions are indicated in red (see section \protect\ref{stab}) and the supersymmetric
solutions in light green. By a suitable rescaling of the coefficients the dependence on the overall scale $a$ is taken out.
}
\label{plotsolSp2}
\end{figure}

Plugging eqs.~\eqref{Sp2geom} into the supergravity equations of motion \eqref{eomconds} we find numerically a rich spectrum of solutions, which are displayed
in figures \ref{plotR4dSp2} and \ref{plotsolSp2}. Note that the dependence on the overall scale can be easily extracted from all plotted quantities
by multiplying by $a$ to a suitable power.
We plotted the value of the 4D Ricci scalar $R_{\text{4D}}$ of the AdS-space against the shape parameter $\sigma$ in
figure \ref{plotR4dSp2}. Note that $R_{\text{4D}}$ is inversely proportional to the AdS-radius squared and
related to the effective 4D cosmological constant and the vev of the 4D scalar potential $V$ as follows
\eq{
\Lambda = \langle V \rangle =R_{\text{4D}}/4  \, .
}
The supersymmetric solutions are plotted in light green, while
red is used for the non-supersymmetric solutions found to be unstable in section \ref{stab}.
For completeness of the presentation of our numeric results, we provide the values of each of the coefficients
$f_i$ of the ansatz \eqref{ansatz} in figure \ref{plotsolSp2}.

The first point
to note is that where the supersymmetric solutions are restricted to the interval $\sigma \in [2/5,2]$, there exist
non-supersymmetric solutions in the somewhat larger interval $\sigma \in [0.39958,2.13327]$. Furthermore, there are up
to five non-supersymmetric solutions for each supersymmetric solution.

We remark that the parameters $\sigma$ and the overall scale are not continuous moduli since they are
determined by the vevs of the fluxes, which in a proper string theory treatment should be quantized. Indeed,
in the next section we will show that generically all moduli are stabilized. We leave the analysis
of flux quantization, which is complicated by the fact that there is non-trivial $H$-flux (twisting the RR-charges),
to further work. The expectation is that the continuous line of supergravity solutions is replaced by discrete solutions.

Let us now take a look at some special values of $\sigma$. For $\sigma=1$ we find five solutions
of which three (including the supersymmetric
one) are up to scaling equivalent to the solutions \eqref{G2sols} on $\frac{\text{G}_2}{\text{SU(3)}}$ of the previous section \cite{romansIIA,gennonsusy,cassred,cassaninonsusy}.
They have $f_3 = f_5 =0$ and so the fluxes are completely expressed in terms of $J$. However, there are also two new non-supersymmetric solutions
(the dark green and the purple one) which have $f_3\neq 0, f_5 \neq 0$.

Next we turn to the case $\sigma=2$. This point is special in that the metric becomes the
Fubini-Study metric on $\mathbb{CP}^3$ and the bosonic symmetry of the geometry enhances from Sp(2) to SU(4).
In fact, since the RR-forms of the supersymmetric solution can be expanded in terms of the closed K\"ahler form
$\tilde{J}=(1/3) J + (2a)^{1/2} W_2$ of the Fubini-Study metric, the symmetry group of the whole supersymmetric solution
is SU(4). One can also show that the supersymmetry enhances from the generic $N=1$ to $N=6$ \cite{nilssonpope}.
In \cite{dimitriosextrasol} it was found that there is an infinite continuous family of non-supersymmetric solutions
and two discrete separate solutions (see also \cite{romansIIA} for an incomplete early discussion),
which all have SU(4)-symmetry. They are {\em not} displayed in the
plot since they can not be found by taking a continuous limit $\sigma \rightarrow 2$.
For these solutions $H=0$ ($f_7=0$) and $\hat{F}_2$ and $\hat{F}_4$ are expanded in terms of $\tilde{J}$ (for more details see
\cite{dimitriosextrasol}).

Instead, in the plot we find apart from the supersymmetric solution (which
merges with the dark green solution at $\sigma=2$) two more discrete non-supersymmetric solutions, which have
only Sp(2)-symmetry (since the fluxes cannot be expressed in terms of $\tilde{J}$ only). The blue one is new, while
the red one turns out to be the reduction
of the Englert-type solution. Indeed for the Englert-type solution we expect
\allowdisplaybreaks\subeq{\al{
& f_1 = 0 \, , & \text{no Romans mass} \, , \\
& f_2 = f_{2,\text{susy}} \, , \quad f_3 = f_{3,\text{susy}} \, , & \text{same geometry in M} \Rightarrow \text{same } \hat{F}_2 \text{ as susy} \, , \\
& f_7 = -2 f_4 = -(1/3) f_{6,\text{susy}} \, , \quad f_5=0 \, ,  & \text{from } \hat{F}_4 \text{ in M-theory} \, , \\
& f_6 = (-2/3) f_{6,\text{susy}} \, , & \text{Freund-Rubin parameter changes} \, , \\
& R_{\text{4D}} = (5/6) R_{\text{4D},\text{susy}} \, , & \text{4D }\Lambda\text{ changes} \, ,
}}
which agrees with the values displayed in the figures for the red curve at $\sigma=2$.

Also for $\sigma=2/5$ we find apart from the supersymmetric solution, the Englert solution (the purple curve)
and one extra non-supersymmetric solution (the dark green curve). Note that while the supersymmetric curve joins the olive green
curve at $\sigma=2/5$, the purple curve only joins the dark green curve at $\sigma=0.39958$.

\subsection*{$\frac{\text{SU(3)}}{\text{U(1)}\times \text{U(1)}}$}

For this manifold the SU(3)-structure is given by \cite{cosets,effectivecosets}:
\eq{\spl{
J & =  a (- e^{12} + \rho e^{34} - \sigma e^{56}) \, , \\
\Omega & = a^{3/2} (\rho \sigma)^{1/2} \left[ (e^{245}+e^{135}+e^{146}-e^{236}) + i (e^{235}+e^{136}+e^{246}-e^{145}) \right] \, ,
}}
where $\rho$ and $\sigma$ are the shape parameters of the model.
Furthermore we find for the torsion classes:
\eq{\spl{
W_1 & = -(a\rho\sigma)^{-1/2} \, \frac{1+\rho+\sigma}{3} \, , \\
W_2 & = -(2/3) a^{1/2} (\rho \sigma)^{-1/2} \left[ (2 - \rho - \sigma) e^{12}
+\rho (1- 2 \rho +\sigma) e^{34} - \sigma (1+\rho -2 \sigma) e^{56} \right] \, .
}}
It turns out that the ansatz \eqref{W2W2} is only satisfied
for
\eq{
\label{rhosigres}
\rho=1 \, , \quad  \sigma=1  \quad \text{or} \quad \rho=\sigma \, .
}
In all three of these cases the equations \eqref{eomconds} for $\frac{\text{SU(3)}}{\text{U(1)}\times \text{U(1)}}$
reduce to exactly the same equations as for $\frac{\text{Sp(2)}}{\text{S}(\text{U(2)}\times \text{U(1)})}$
so that we obtain the same solution space. However, as we will see in the next section, the stability analysis
will be different since the model on $\frac{\text{SU(3)}}{\text{U(1)}\times \text{U(1)}}$ has two extra left-invariant modes.

In order to find further non-supersymmetric solutions, we should go beyond
the ansatz \eqref{W2W2}. Let us put
\eq{
\hat{W}_2 \wedge \hat{W}_2 = (-1/6) \, J \wedge J + p_1 \, \hat{W}_2 \wedge J + p_2 \hat{P} \wedge J \, ,
}
where $\hat{P}$ is a primitive normalized (1,1)-form (so that it is orthogonal to $J$ and $\hat{P}^2=1$). Furthermore, we
also choose it orthogonal to $\hat{W}_2$ i.e.
\eq{
\hat{W}_2 \cdot \hat{P} = 0 \, \quad \text{or equivalently} \quad J \wedge \hat{W}_2 \wedge \hat{P} = 0 \, .
}
{} From the last equation one finds, using \eqref{imomder}, that $\d \hat{P} \wedge \Im \Omega=0$, which implies
on $\frac{\text{SU(3)}}{\text{U(1)}\times \text{U(1)}}$ that
\eq{
\d \hat{P} = 0 \, .
}
One can now allow the RR-fluxes $\hat{F}_2$ and $\hat{F}_4$ to have pieces
proportional to $\hat{P}$ and $\hat{P} \wedge J$ respectively and adapt the equations \eqref{eomconds}
accordingly to accommodate for the new contributions. Now it is possible
to numerically find non-supersymmetric solutions for $\rho$ and $\sigma$ not satisfying \eqref{rhosigres}.
In particular, there are Englert-type solutions on the ellipse of values for $(\rho,\sigma)$ where the supersymmetric
solution has zero Romans mass. From eq.~\eqref{romanssusy} we find that this ellipse is described by
\eq{
m^2 = \frac{5}{16 \rho \sigma} \left[ -5 (\rho^2+\sigma^2) + 6 (\rho + \sigma + \rho\sigma) -5 \right] = 0 \, .
}
We will not go into more detail on these solutions in this paper.

\section{Stability analysis}\label{stab}

In this section we investigate whether the new non-supersymmetric solutions on $\frac{\text{Sp(2)}}{\text{S}(\text{U(2)}\times \text{U(1)})}$
and $\frac{\text{SU(3)}}{\text{U(1)}\times \text{U(1)}}$ are stable\footnote{In \cite{cassred} it was found
that the non-supersymmetric solutions on $\frac{\text{G}_2}{\text{SU(3)}}$ and the similar solutions on the nearly-K\"ahler
limits of the other two coset manifolds under study are stable. We find exactly the same spectrum as the authors of that paper, which provides
a consistency check on our approach. We thank Davide Cassani for providing us with these numbers, which are not explicitly given in their paper.
We did not investigate the spectrum of the similar solution on $\text{SU(2)}\times\text{SU(2)}$, which is more complicated as there
are more modes.}. To this end we calculate
the spectrum of scalar fluctuations. We use the well-known result of \cite{bf1,bf2} that in an AdS$_4$ vacuum a
tachyonic mode does not yet signal an instability. Only a mode with a mass-squared below the Breitenlohner-Freedman
bound,
\eq{
\label{bfbound}
M^2 < -\frac{3|\Lambda|}{4} \, ,
}
where $\Lambda < 0$ is the 4D effective cosmological constant,
leads to an instability. We restrict ourselves to left-invariant fluctuations, which implies that even if we do not find any
modes below the Breitenlohner-Freedman bound, the vacuum might still be unstable, since there might be fluctuations with sufficiently
negative mass-squared that are not left-invariant. This analysis can however pinpoint many unstable vacua and we do believe
it gives a valuable first indication for the stability of the others.

Truncating to the left-invariant modes on the coset manifolds under study leads to a 4D
$N=2$ gauged supergravity\footnote{It is important to make the distinction between the number of supersymmetries
of respectively the 4D effective theory, the 10D compactifications, and their 4D truncation (which are the solutions of the 4D effective theory \cite{cassred}). In the presence of one left-invariant internal spinor, the effective theory will
be $N=2$ since this same spinor can be used in the $4+6$ decomposition of both ten-dimensional Majorana-Weyl supersymmetry
generators, but multiplied with independent four-dimensional spinors. On the other hand,
for a certain compactification to preserve the supersymmetry, certain differential conditions, which follow from
putting the variations of the fermions to zero must be satisfied. In the presence of RR-fluxes, these conditions mix both ten-dimensional
Majorana-Weyl spinors, putting the four-dimensional spinors in both decompositions equal. A generic supersymmetric compactification therefore only preserves $N=1$.
The $\sigma=2$ supersymmetric K\"ahler-Einstein solution on $\mathbb{CP}^3$ on the other hand is non-generic in that it preserves $N=6$, of which only one internal spinor
is left-invariant under the action of Sp(2) and remains after truncation to 4D.}.
It has been shown in \cite{cassred} that this truncation
is consistent. The spectrum of the scalar fields can then be obtained from the 4D scalar potential.
In fact, this computation is analogous to the one performed in \cite{effectivecosets} for the
supersymmetric $N=1$ vacua on the coset spaces. As opposed to the models here, the models in that paper
included orientifolds, which broke the supersymmetry of the 4D effective theory from $N=2$ to $N=1$.
However, also in the present case the $N=1$ approach is applicable and effectively we have used exactly
the same procedure, i.e.\ using the $N=1$ scalar fluctuations and obtaining the scalar potential
from the $N=1$ superpotential and K\"ahler potential (see \cite{lg1,lg2,grimmN1,effective}).\footnote{It 
is interesting to note that (in $N=1$ language) all the D-terms vanish, so that
the supersymmetry breaking is purely due to F-terms. Indeed,
in \cite{effective} it is shown that $\mathcal{D}=0$ is equivalent to $\d_H(e^{2A-\Phi} \Re \Psi_1)=0$ in the
generalized geometry formalism. For SU(3)-structure this translates to $\d (e^{{2A}-\Phi} \Re \Omega)=0$ and $H \wedge \Re \Omega=0$,
which is satisfied for our ansatz, eq.~\eqref{dJdOm} and \eqref{ansatz}.}
The reason is the following. The $N=2$ scalar fluctuations in the vector multiplets are
\eq{
J_c = J - i  B = (k^i - i b^i) \omega_i = t^i \omega_i \, ,
}
where $\omega_i$ span the left-invariant two-forms of the coset manifold. The orientifold projection of the $N=1$ theory would then project
out the scalar fluctuations
coming from expanding on {\em even} two-forms, which are absent for the $N=1$ theory on the coset manifolds under study. The scalar fluctuations
in the $N=2$ vector multiplets are thus exactly the same as the scalars in the chiral multiplets of the K\"ahler moduli sector of
the $N=1$ theory. The expansion forms can then be chosen to be the same as the $Y^{(2-)}_i$ of \cite{effectivecosets}.
Furthermore, there is one  tensor multiplet, which contains the dilaton $\Phi$, the two-form $B_{\mu\nu}$ and two axions $\xi$ and $\tilde{\xi}$ coming from the expansion
of the RR-potential $C_3$:
\eq{
C_3 = \xi \, \alpha + \tilde{\xi} \, \beta \, ,
}
where a choice for $\alpha$ and $\beta$ spanning the left-invariant three-forms would be
$Y^{(3-)}$ and $Y^{(3+)}$ of \cite{effectivecosets} respectively. In the presence of Romans mass
or $\hat{F}_2$-flux the two-form $B_{\mu\nu}$ becomes massive and cannot be dualized to a scalar.
The dilaton and $\tilde{\xi}$ appear in a chiral multiplet of the complex moduli sector of the $N=1$ theory, while $B_{\mu\nu}$
and $\xi$ are projected out by the orientifold. By using the $N=1$ approach we thus loose
the information on just one scalar $\xi$. A proper $N=2$ analysis would however learn that
$\xi$ does not appear in the scalar potential (see e.g.~\cite{cassred}), implying that it is massless
and thus above the Breitenlohner-Freedman bound.
Moreover, the scalar potential should be the same whether it is obtained directly from reducing the 10D
supergravity action (as in \cite{casspot}) or whether it is obtained using $N=2$ or $N=1$ technology\footnote{The
only potential difference between the latter two would be the contribution from the orientifold. We
have checked that this contribution vanishes in the scalar potentials of \cite{effectivecosets} in the limit
of the orientifold charge $\mu \rightarrow 0$.}.
Furthermore we note that the massless scalar field $\xi$ not appearing in the potential is not a modulus,
since it is charged \cite{louismicu,louismirror}, and therefore eaten by a vector field becoming massive.

\begin{figure}
\centering
\subfigure[Spectrum of $\frac{\text{Sp(2)}}{\text{S}(\text{U(2)}\times \text{U(1)})}$]{
\includegraphics[width=7cm]{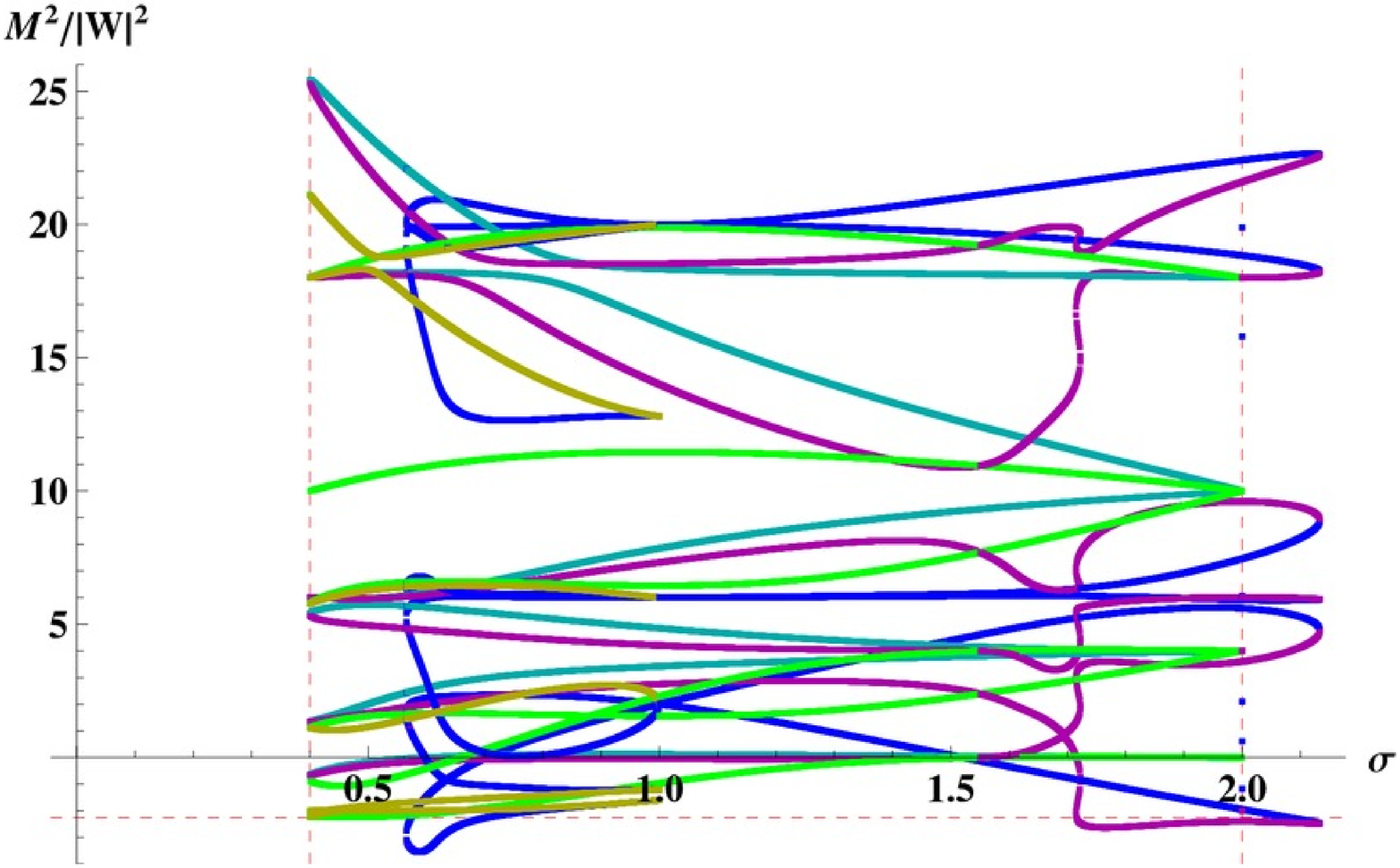}
}
\hspace{0.2cm}
\subfigure[Two extra modes of the $\frac{\text{SU(3)}}{\text{U(1)}\times \text{U(1)}}$-model]{
\includegraphics[width=7cm]{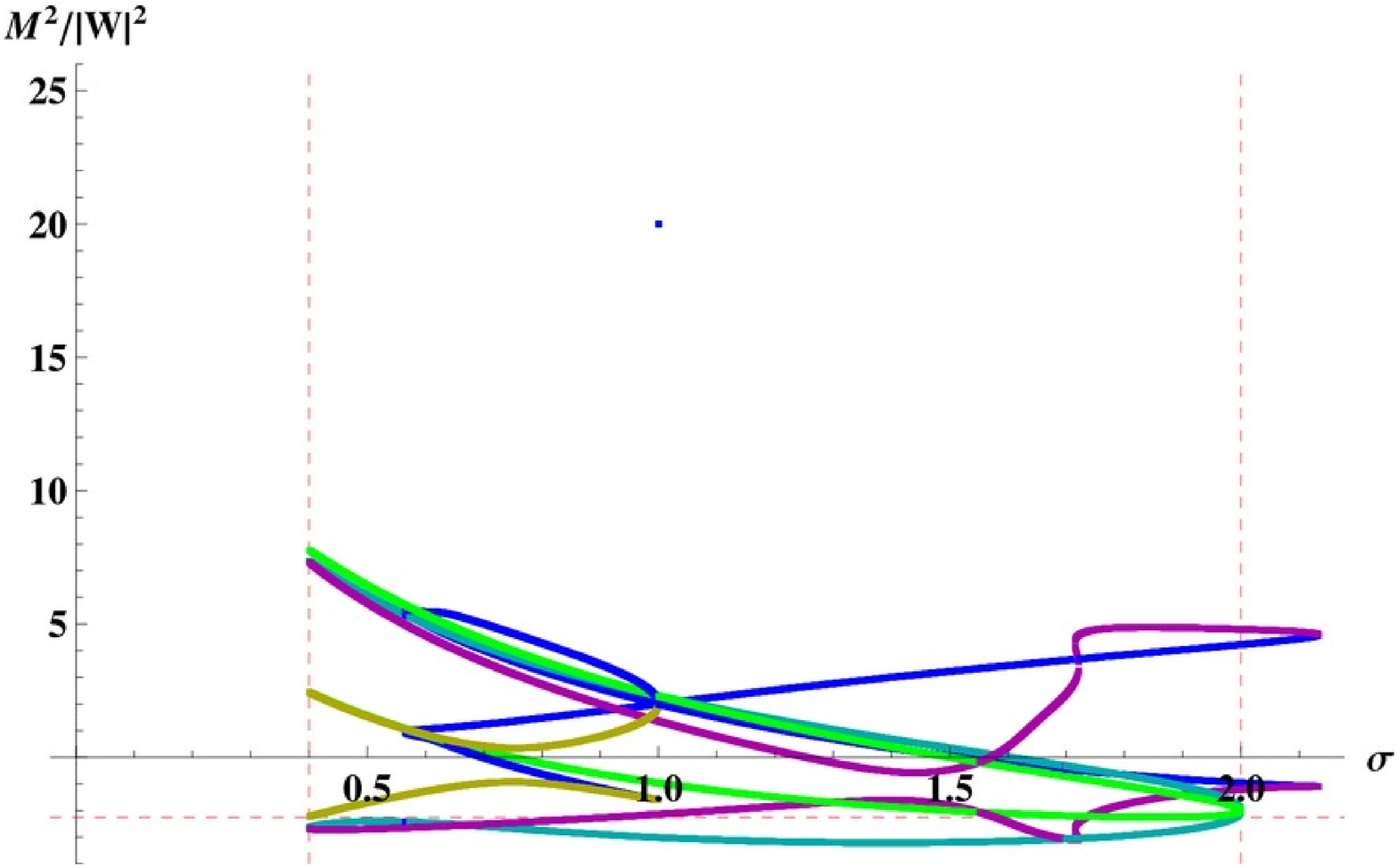}
}
\caption{Spectrum of left-invariant modes of the solutions on
$\frac{\text{Sp(2)}}{\text{S}(\text{U(2)}\times \text{U(1)})}$ and $\frac{\text{SU(3)}}{\text{U(1)}\times \text{U(1)}}$.}
\label{spectrum}
\end{figure}

\begin{figure}
\centering
\includegraphics[width=14cm]{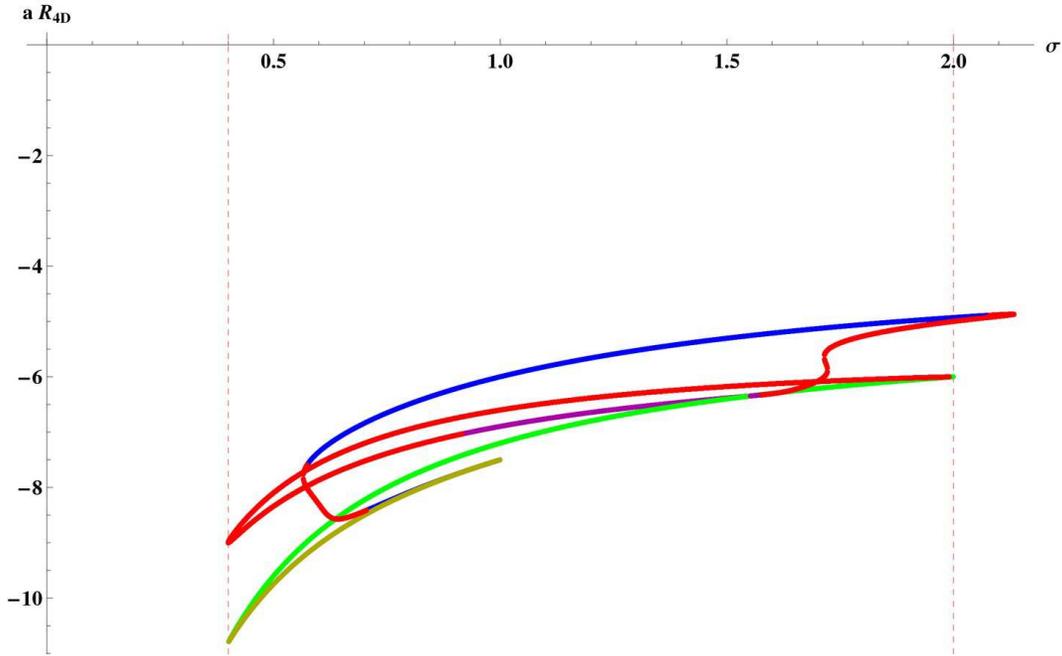}
\caption{$\frac{\text{SU(3)}}{\text{U(1)}\times \text{U(1)}}$-model: plot of $a R_{\text{4D}}$ in terms of the shape parameter $\sigma$.
Unstable solutions are indicated in red.}
\label{plotR4dSU3}
\end{figure}

The spectra of left-invariant modes for $\frac{\text{Sp(2)}}{\text{S}(\text{U(2)}\times \text{U(1)})}$
and $\frac{\text{SU(3)}}{\text{U(1)}\times \text{U(1)}}$ are displayed in figure \ref{spectrum}. The
Breitenlohner-Freedman bound is indicated as a horizontal dashed line. The Sp(2)-model has six scalar fluctuations entering the
potential: $k^i,b^i$ with $i=1,2$ from the two vector multiplets, and $\Phi,\tilde{\xi}$ from the universal
hypermultiplet, while the SU(3)-model has two more fluctuations from the extra vector multiplet.
These two extra modes make a big difference for the stability analysis since one of them tends to be below the Breitenlohner-Freedman
bound for the purple and dark green solution. As a result, even though the solutions for the Sp(2)- and SU(3)-model take the same
form, the SU(3)-model has more unstable solutions: compare figure \ref{plotR4dSp2} and \ref{plotR4dSU3}.

In particular, we note that the reduction of the Englert-type solution is unstable for $\sigma=2$ in the Sp(2)-model, in agreement
with \cite{englertunstable}, since the M-theory lift of the corresponding supersymmetric solution has eight Killing spinors.
We indeed find the same negative mass-squared $M^2 = -(4/5) |\Lambda|$ for the unstable mode
as in that paper. On the other hand, for $\sigma=2/5$ the Englert-type solution is stable against left-invariant fluctuations.
This is still in agreement with \cite{englertunstable} which relied on the existence of at least two Killing spinors, while the M-theory lift of
the $N=1$ supersymmetric solution at $\sigma=2/5$ has only one Killing-spinor. For the SU(3)-model, all Englert-type solutions turn out to be unstable (including
the ones outside the condition \eqref{rhosigres}).

We also investigated the stability of the additional solutions at the special point $\sigma=2$ found in \cite{dimitriosextrasol}.
We found that for the Sp(2)-model all these solutions are stable against left-invariant fluctuations. For the SU(3)-model
on the other hand it turns out that the discrete solutions in eqs.~(3.16) and (3.17) of that reference are unstable, while
the continuous family of eq.~(3.18) becomes unstable for
\eq{
\frac{\gamma^2}{\beta^2} > \frac{5(75\mp 16 \sqrt{21})}{8217} \, ,
}
for the $\pm$ sign choice in front of the square root in eq.~(3.18) of that paper respectively (note that the supersymmetric solution
corresponds to the point $\gamma^2/\beta^2=0$ in this family).

Finally, we note that generically (i.e.\ unless an eigenvalue is crossing zero at a special value for $\sigma$)
all the plotted modes are massive. For a range of values for $\sigma$ one of the eigenvalues for the dark green and purple
solution takes a small, but still non-zero value.

\section{Conclusions}

In this paper we presented new families of non-supersymmetric AdS$_4$ vacua. In fact, extrapolating from our analysis
on these specific coset manifolds and under the assumption that a proper treatment of flux quantization does not kill
much more vacua than in the supersymmetric case, it would seem that there are more of these non-supersymmetric vacua than supersymmetric
ones. This would imply that such vacua cannot be ignored in landscape studies. We have moreover shown that many of them are stable against a specific set of fluctuations, namely the ones that can be expanded in terms of left-invariant forms. If these vacua turn out to be stable
against all fluctuations they should also have a CFT-dual, which could be studied along the lines of \cite{tomasiellomassive},
where the three-dimensional Chern-Simons-matter theory dual to a particular highly symmetric non-supersymmetric vacuum was
proposed. Furthermore, the nice property of some IIA vacua that all moduli enter the superpotential and thus can be stabilized at a classical level \cite{dewolfe} also extends to our non-supersymmetric vacua.

A next step would be to relax the constraint that the solutions should have the same geometry as the supersymmetric solution.
It is also interesting to investigate whether a similar ansatz and techniques can be used to look for tree-level dS-vacua \cite{upssala}.

\begin{acknowledgments}
We thank Davide Cassani for useful email correspondence and proofreading, and furthermore Claudio Caviezel for
active discussions and initial collaboration. We would further like to thank the Max-Planck-Institut f\"ur Physik in Munich,
where both of the authors were affiliated during the bulk of the work on
this paper. P.K.\ is a Postdoctoral Fellow of the FWO -- Vlaanderen.
The work of P.K.\ is further supported in part by the FWO -- Vlaanderen project
G.0235.05 and in part by the Federal Office for Scientific, Technical and
Cultural Affairs through the 'Interuniversity Attraction Poles
Programme Belgian Science Policy' P6/11-P. S.K.\ is supported by the SFB -- Transregio 33 ``The Dark Universe'' by
the DFG.
\end{acknowledgments}

\appendix

\section{SU(3)-structure}\label{asu3}

A real non-degenerate two-form $J$ and a complex decomposable
three-form $\Omega$ define an SU(3)-structure on the 6D manifold $M_6$ iff:
\begin{subequations}\label{OmegaJ}
\begin{align}
\Omega\wedge J&=0 \, , \\
\Omega\wedge\bar{\Omega}&=\frac{8i}{3!}J\wedge J \wedge J \neq 0 \, ,
\label{a1}
\end{align}
\end{subequations}
and the associated metric is positive-definite. This metric
is determined by $J$ and $\Omega$ as follows:
\eq{\label{su3metric}
g_{mn}=-J_{mp}I^p{}_n \, ,
}
with $I$ the complex structure associated (in the way of \cite{hitchinold}) to $\Omega$.
The volume-form is given by $\text{vol}_6 = \frac{1}{3!}J^3 = -(i/8) \Omega \wedge \bar{\Omega}$.

The intrinsic torsion of the manifold $M_6$ decomposes into five torsion classes
${\cal W}_1,\dots,{\cal W}_5$. Alternatively they correspond to the
SU(3)-decomposition of the exterior derivatives of $J$ and $\Omega$ \cite{chiossal}. Intuitively,
they parameterize the failure of the manifold to be of special holonomy, which can also
be thought of as the deviation from closure of $J$ and $\Omega$.
More specifically we have:
\eq{\spl{
\d J&=\frac{3}{2}\Im(\mathcal{W}_1\bar{\Omega})+\mathcal{W}_4\wedge J+\mathcal{W}_3 \, , \\
\d \Omega&= \mathcal{W}_1 J\wedge J+\mathcal{W}_2 \wedge J+\bar{\mathcal{W}}_5\wedge \Omega ~,
\label{torsionclasses}
}}
where $\mathcal{W}_1$ is a scalar, $\mathcal{W}_2$ is a primitive (1,1)-form, $\mathcal{W}_3$ is a real
primitive $(1,2)+(2,1)$-form, $\mathcal{W}_4$ is a real one-form and $\mathcal{W}_5$ a complex (1,0)-form.
In this paper only the torsion classes $\mathcal{W}_1$, $\mathcal{W}_2$ are non-vanishing and they are
purely imaginary, so it will be convenient to define $W_{1,2}$ so that $\mathcal{W}_{1,2}=i W_{1,2}$.
A primitive (1,1)-form $P$ (such as $W_2$) transforms under the $\bf{8}$ of SU(3) and satisfies
\eq{
\label{primitive}
P \wedge J \wedge J = 0 \, .
}
The Hodge dual is given by
\eq{
\label{simplehodge}
\star_6 P = - P \wedge J \, .
}
A primitive $(1,2)$(or $(2,1)$)-form $Q$ on the other hand transforms as a $\bf{6}$ (or $\bar{\bf{6}}$) under SU(3) and satisfies
\eq{
Q \wedge J = 0 \, .
}

\section{Type II supergravity}
\label{asugra}

The bosonic content of type II supergravity consists of a metric $G$, a dilaton $\Phi$, an NSNS three-form $H$ and RR-fields
$F_{n}$. We use the democratic formalism of \cite{democratic}, in which
the number  of  RR-fields is doubled, so that $n$ runs over $0,2,4,6,8,10$ in type IIA and over $1,3,5,7,9$ in IIB.
We will often collectively denote the RR-fields with the polyform $F=\sum_n F_{n}$. We have also doubled
the RR-potentials, collectively denoted by $C=\sum_n C_{(n-1)}$. These potentials
satisfy $F=\d_H C + m e^{-B}=(\d + H\wedge)C + m e^{-B}$. In type IIB there is of course no Romans mass $m$, so
that the second term vanishes. In type IIA we find in particular $F_0=m$.

The bosonic part of the pseudo-action of the democratic formalism then simply reads
\eq{
\label{pa}
S = \frac{1}{2 \kappa_{10}^2} \int d^{10} X \sqrt{-G} \left\{ e^{-2 \Phi} \left[ R + 4 (\d \Phi)^2 -\frac{1}{2} H^2 \right] - \frac{1}{4} F^2 \right\} \, ,
}
where we defined $F^2 = \sum_n F_{n}^2$ and the square of an $l$-form $P$ as follows
\subeq{\label{formsquared}
\eq{
P^2 = P \cdot P = \frac{1}{l!} P_{m_1 \ldots m_l} P^{m_1 \ldots m_l} \, ,
}
where the indices are raised with the inverse of the metric $G_{mn}$ or the internal metric $g_{mn}$ (defined
later on), depending on the context. In the following it
will also be convenient to define:
\eq{
P_m \cdot P_n = \iota_m P \cdot \iota_n P = \frac{1}{(l-1)!} P_{mm_2 \ldots m_l} P_n{}^{m_2 \ldots m_l} \, .
}}
The extra degrees of freedom for the RR-fields in the democratic formalism
have to be removed by hand by imposing the following duality condition at the level of the equations
of motion after deriving them from the action \eqref{pa}:
\eq{
\label{Fduality}
F_{n} = (-1)^{\frac{(n-1)(n-2)}{2}} \star_{10} F_{10-n} \, .
}
That is why \eqref{pa} is only a pseudo-action.

The fermionic content consists
of a doublet of gravitinos $\psi_{M}$ and a doublet of dilatinos $\lambda$.
The components of the doublets are of different chirality
in type IIA and of the same chirality in type IIB.

In this paper we look for vacuum solutions that take the form AdS$_4 \times M_6$.
In principle there could also be a warp factor $A$, but it will always be constant for the solutions
in this paper. We can choose it to be zero. The compactification ansatz for the metric then reads
\eq{
\d s_{10}^2 = G_{mn} \d X^m \d X^n=\d s_{4}^2 + g_{mn} \d x^m \d x^n \, ,
}
where $\d s_{4}^2$ is the line-element for AdS$_4$ and $g_{mn}$ is the metric on the internal
space $M_6$. For the RR-fluxes the ansatz becomes
\eq{
\label{fluxcomp}
F = \hat{F} + \text{vol}_4 \wedge \tilde{F} \, ,
}
where $\hat{F}$ and $\tilde{F}$ only have internal indices. The duality constraint \eqref{Fduality}
implies that $\tilde{F}$ is not independent of $\hat{F}$, and given by
\eq{
\label{Fduality6d}
\tilde{F}_{n} = (-1)^{\frac{(n-1)(n-2)}{2}} \star_6 \hat{F}_{6-n} \, .
}

What we need in this paper are the type II equations of motion, which can be found from the pseudo-action
\eqref{pa}. We use them as they are written down in \cite{integr} (originally they were obtained for massive type IIA
in \cite{romansIIA}), but take some linear combinations in order to further
simplify then. Without source terms (i.e.~we put $j_{\text{total}}=0$ in the equations of motion of \cite{integr}),
they then read:
\allowdisplaybreaks
\subeq{\label{sugraeom}\al{
& \d_H F = 0 && (\text{Bianchi RR fields}) \, ,  \\
& \d_{-H} \star_{10} F = 0 &&(\text{eom RR fields}) \, , \\
& \d H =0 &&(\text{Bianchi }H) \, , \\
& \d \left( e^{-2\Phi} \star_{10} H \right) - \frac{1}{2} \sum_n \star_{10} F_{n} \wedge F_{n-2} = 0 &&(\text{eom }H) \, , \\
& 2 R - H^2 + 8 \left(\nabla^2 \Phi - (\partial \Phi)^2 \right) = 0 &&(\text{dilaton eom}) \, , \\
& 2(\partial \Phi)^2 - \nabla^2 \Phi -\frac{1}{2} H^2 - \frac{e^{2\Phi}}{8} \sum_n n F_{n}^2 = 0
&& (\text{trace Einstein/dilaton eom}) \, , \\
& R_{MN} + 2 \nabla_M \partial_N \Phi - \frac{1}{2} H_M \cdot H_N - \frac{e^{2\Phi}}{4} \sum_n F_{n\,M} \cdot F_{n\,N} =0  \hspace{-2cm} \\
&&& (\text{Einstein eq./dilaton/trace}) \, . \nonumber
}}


\bibliography{nonsusy}

\end{document}